\documentclass{llncs}
\usepackage{mathptmx}       
\usepackage{helvet}         
\usepackage{courier}        
\usepackage{type1cm}        
\usepackage{makeidx}         
\usepackage{graphicx}        
\usepackage{multicol}        
\usepackage[bottom]{footmisc}
\usepackage{subfigure}
\usepackage{amsfonts}
\usepackage[cmex10]{amsmath}
\usepackage{cite}
\usepackage{caption}
\begin{document}

\title{Detecting People Interested in Non-Suicidal Self-Injury on Social Media}

%
%
\author{Zaihan Yang \and Dmitry Zinoviev}
%
%
%
\institute{Suffolk University, Boston, MA 02108, USA\\
  \email{zyang13@suffolk.edu}, \email{dzinoviev@suffolk.edu}}

\maketitle    

\section{Introduction}
Non-Suicidal Self-Injury (NSSI) is the intentional destruction of body tissue without the intent to commit suicide~\cite{Apa}.  It is particularly prevalent among adolescents and young adults as a means of emotional control and release. Typical NSSI activities include skin cutting, banging or hitting oneself, and burns.

Recent prevalence estimates suggest that 14\%--21\% of adolescents and 17\%--25\% of young adults have engaged in NSSI at some point in their lives. NSSI is repeatedly found to be associated with significant emotional and behavioral dysfunction (such as eating disorders and suicide). This relationship highlights the urgency of providing early detection of people with NSSI engagement and prevention of their behaviors.

However, global provisions and services for detecting, supporting, and treating NSSI people have long been insufficient.  There is no reliable laboratory test for diagnosing NSSI. Diagnostic largely depends on patients' self-reports or observations reported by relatives or friends. Yet, NSSI people often conceal their practices, which prevents detecting their engagement.  

Early research work on NSSI people detection was primarily conducted within psychology, psychiatry, and medicine domains~\cite{Apa, Favazza, Andover}. With the proliferation of social media, people are increasingly using online platforms to share their thoughts and opinions.  Postings on these sites are made in natural settings and provide a means for capturing people's real thoughts, opinions, and moods. Researchers from Computer and Data Science fields have started to explore social media content to study people with NSSI engagement, their interests~\cite{Zinoviev}, the influence of social media on their behaviors~\cite{Daine}, and their posted images~\cite{Xian}. However, to the best of our knowledge, no work has been done to provide an automatic learning system that can detect people with NSSI engagement.

We treat the detection of people interested in NSSI as a binary classification problem. We have collected data from LiveJournal.com, a social-blogging networking platform, and built Na\"{\i}ve Bayes and Logistic Regression classifiers based on the features extracted from users' self-declared interests. Experimental evaluation demonstrates that we can achieve 73\% accuracy, 77\% precision, 67\% recall, and 71\% F$_1$ score to detect people interested in NSSI and identify the most discriminating features.

\section{Model}
We have collected our data by crawling user profiles (demographics, self-declared interests, and friendship relationships) and NSSI-related thematic community profiles (membership, posts, and comments).  

We assume that any LiveJournal user who is a member of one of the 139 manually selected NSSI-related thematic communities or contributes to such community by posting or commenting, is interested in NSSI. We designate such users as ``harmers'' for brevity, acknowledging that some may not practice self-injury. Following the harmers' friendship network, we further collect some of their immediate friends and friends-of-friends (the ``non-harmers''), chosen randomly to match the size and age of the harmers.

The dataset has 11,972 harmers, 12,600 friends, 11,672 friends-of-friends, and 1,264 distinct self-declared single- or multi-word interests (on average, 26.2 interests per user). These self-declared interests serve as virtual profiles for each user. 
We regard each interest as a feature and represent the feature vectors in the following three ways:
\begin{description}
\item[Simple-Count-Vector:] The feature value of each interest is represented by their occurrence frequency in user profiles.
\item[TF-IDF-Vector:] Each feature is represented by its TF-IDF value.
\item[Topic-Distribution-Vector:] We apply the Latent Dirichlet Allocation (LDA) model to learn a topic distribution over each user profile. We use the distribution as the feature vector. We choose the number of topics to be 10.
\end{description}


\section{Results}
We treat the harmers as positive samples and the non-harmers as negative samples. 
Since the original data set is unbalanced (with 1/3 positive and 2/3 negative samples), we pick 11,972 negative samples uniformly at random to form a balanced data set, and only consider those interests that are declared by more than 100 but fewer than 16,760 (70\% of all users in the balanced data set) users to construct the three feature vectors mentioned above. We repeat the random sampling process five times and apply 5-fold cross-validation to evaluate the model's performance. Table~\ref{Table:classification} shows the classification results evaluated by accuracy, precision, recall, and F$_1$ score using the three feature vector representations and Na\"{\i}ve Bayes (NB) and Logistic Regression (LR) classifiers.

\begin{table}[ht]
\centering
\caption{\label{Table:classification}Classification results \\ \small{(Highest value of each metric is highlighted in bold)}}
\begin{tabular}{|l|c|c|c|c|}
\hline 
Feature Vector (Classifier) & Accuracy & Precision & Recall & F$_1$ Score \\ 
\hline 
Simple-Count (NB) & 0.70 & 0.76 & 0.59 & 0.66 \\ 
\hline 
TF-IDF (NB) & 0.71  & 0.74 & 0.64 & 0.68 \\ 
\hline 
LDA-Topic-Distribution (NB) & 0.65 & 0.67 & 0.60 & 0.63 \\ 
\hline 
Simple-Count (LR) & 0.72 & \textbf{0.77} & 0.62 & 0.69 \\ 
\hline 
TF-IDF (LR) & \textbf{0.73}  & 0.75 & \textbf{0.67} & \textbf{0.71} \\ 
\hline 
LDA-Topic-Distribution (LR) & 0.67 & 0.72 & 0.56 & 0.63 \\ 
\hline 
\end{tabular} 
\end{table}

We achieved up to 73\% accuracy, 77\% precision, 67\% recall, and 71\% F$_1$ score. For both Na\"{\i}ve Bayes and Logistic Regression classifiers, using TF-IDF feature vectors increases recall and F$_1$ score but slightly decreases precision as compared to Simple-Count vectors. Since our project aims to detect concealed behavior, higher recall and false positives are more desirable than a higher precision and false negatives.

We compute and sort each feature by its odds-ratio value learned from Na\"{\i}ve Bayes and report the top 20 positive and negative features. Table~\ref{Table:features} shows the top 20 positive interests prevalent among the harmers (such as cutting, burning, self-injury, and self-harm) and the bottom 20 interests that the harmers tend to neglect (such as programming, linux, architecture, and chess). The harmers' interests are undoubtedly specific to NSSI, related disorders, and mental health in general. The non-harmers interests seem to be aligned with an adolescent/young adult's interests in the early XXI$^\mathrm{st}$ century.

\begin{table}[ht]
\centering
\caption{\label{Table:features}Discriminating features for Simple-Count and TF-IDF}
\begin{tabular}{|p{0.1\textwidth}|p{0.45\textwidth}|p{0.41\textwidth}|}
  \hline
  &Top positive & Top negative \\
  \hline 
  Simple-Count & abuse, ednos, ocd & robots, technology, they might be giants, tom waits, firefly\\\hline 
  TF-IDF & mental health, scars, mental illness & smallville, travel, naruto, american eagle, torchwood\\\hline 
  Both &anxiety, pills, bulmima, suicide, bleeding, hurt, borderline personality disorder, razorblades, bipolar, cut, razors, burning, cutting, self mutilation, self injury, self harm, si & programming, battlestar galactica, linux, d\&d, fanfic, farscape, science fiction, architecture, fishing, animation, tolkien, gaming, chess, rpg, doctor who\\
\hline
\end{tabular}
\end{table}

%
%

\paragraph{Summary.}
In this paper, we propose  a supervised learning approach to detect people interested in NSSI. Experimental evaluation on a real-world dataset---the LiveJournal social blogging networking platform---demonstrates our proposed model's effectiveness.  For future work, in addition to the pure content-based features, we would like to integrate network, demographic, sentimental, contextual, and other features.  We consider building and comparing classifiers based on different algorithms (including logistic regression, random forest, and artificial neural networks). We plan to apply our approach to other social media platforms such as Twitter or Facebook.


\begin{thebibliography}{10}
\providecommand{\url}[1]{\texttt{#1}}
\providecommand{\urlprefix}{URL }

\bibitem{Apa}
American Psychastic Association: Diagnostic and Statistical Manual of Mental Disorders (5th ed.). American Psychiatric Association, 2013

\bibitem{Favazza}
Favazza, A.R.: Bodies under Siege: Self-mutilation, Nonsuicidal Self-Injury, and Body Modification in Culture and Psychiatry (3rd ed.). Johns Hopkins University Press, 2011.

\bibitem{Andover}
Andover, M.: Non-Suicidal Self-Injury Disorder in a Community Sample of Adults. Psychiatry Research. 219, 2014

\bibitem{Zinoviev}
Zinoviev, D., Stefanescu, D., Fireman, G., Swenson, L.: Semantic Networks of Interests in Online Non-Suicidal Self-Injury Communities. Digital Health. Vol.2, Page 1--14, 2016.

\bibitem{Xian}
Xian, L., Vickers, S.D., Giordano, A.L., Lee, J., Kim, I.K., Ramaswamy, L.: \#Selfharm on Instagram: Quantitative Analysis and Classification of Non-Suicidal Self-Injury.  International Conference on Cognitive Machine Intelligence (CogMI). Page 61--70, 2019.

\bibitem{Daine}
Diane, K., Hawton, K., Singaravelu, V., Stewart, A., Simkin, S., Montgomery, P.: The Power of the Web: A Systematic Review of Studies of the Influence of the Internet on Self-Harm and Suicide in Young People.  PLOS. Volume 8, Issue 10, 2013.




\end{thebibliography}
\end{document}